\documentclass[conference]{IEEEtran}
\IEEEoverridecommandlockouts

\usepackage{cite}
\usepackage{amsmath,amssymb,amsfonts}
\usepackage{algorithmic}
\usepackage{graphicx}
\usepackage{textcomp}
\usepackage{xcolor}
\usepackage{booktabs} 
\usepackage{color,soul}
\usepackage{tabularx,hhline}
\usepackage{multirow}
\usepackage{hyperref}
\usepackage{tikz}
\usepackage{textcomp}
\usepackage{enumitem}

\newtheorem{definition}{{Definition}}
\usepackage[strings]{underscore}

\begin{document}

\title{Physics-Informed Neural Networks for Securing Water Distribution Systems}

\author{
\IEEEauthorblockN{\textbf{Solon Falas}\IEEEauthorrefmark{1}, \textbf{Charalambos Konstantinou}\IEEEauthorrefmark{2} \textbf{Maria K. Michael}\IEEEauthorrefmark{1}}
\IEEEauthorblockA{\IEEEauthorrefmark{1}Dept. of Electrical and Computer Engineering, KIOS Research and Innovation Centre of Excellence, University of Cyprus
\IEEEauthorblockA{\IEEEauthorrefmark{2}FAMU-FSU College of Engineering, Center for Advanced Power Systems, Florida State University}
E-mail: sfalas01@ucy.ac.cy, ckonstantinou@fsu.edu, mmichael@ucy.ac.cy}
}
\IEEEaftertitletext{\vspace{-1\baselineskip}}

\IEEEoverridecommandlockouts

\maketitle

\begin{abstract}
Physics-informed neural networks (PINNs) is an emerging category of neural networks which can be trained to solve supervised learning tasks while taking into consideration given laws of physics described by general nonlinear partial differential equations. PINNs demonstrate promising characteristics such as performance and accuracy using minimal amount of data for training, utilized to accurately represent the physical properties of a system's dynamic environment. In this work, we employ the emerging paradigm of PINNs to demonstrate their potential in enhancing the security of intelligent cyberphysical systems. In particular, we present a proof-of-concept scenario using the use case of water distribution networks, which involves an attack on a controller in charge of regulating a liquid pump through liquid flow sensor measurements. PINNs are used to mitigate the effects of the attack while demonstrating the applicability and challenges of the approach.
\end{abstract}


\vspace{-0.3em}
\section{Introduction} \label{s:Introduction}
Cyberphysical systems (CPSs) integrate physical processes with automation, computation, and computer networks. The cyber environment monitors and controls the behavior of the physical system based on informed decision making mechanisms through feedback loops, usually comprised of a variety of sensors and actuators. Examples of CPSs include water management and distribution systems, power grids, and industrial manufacturing systems. In smart water distribution systems, the deployed sensors monitor vital information such as water level, pressure, and velocity in pipes and provide the system operator an overview of the system's state \cite{rasekh2016smart}. Actuators, such as pumps and flow regulators/valves, act upon the aforementioned measurements to ensure stable and efficient operation. While infusing dynamics and physical processes with software and networking provides many benefits to CPS infrastructures, it also makes them vulnerable to cyberattacks.

The security of CPSs, such as those deployed in critical infrastructures, is of crucial importance due to the catastrophic consequences that may occur in case of system failures. Embedded systems for CPS monitoring and control, spanning over large areas, are likely to be controlled over wireless networks. Despite the benefits of remote control over geographically dispersed locations, such configurations might provide the opportunity to attackers to gain access to the network. As a result, they can maliciously tamper sensor data and disrupt the normal operating conditions of the system.

Recent cyberattacks on CPS infrastructure portray the prevalence and importance of such incidents. For example, on April 24 2020, an event was reported by Israeli officials stating that a cyberattack, aimed at disrupting water supplies by increasing chlorine levels in drinking water, was thwarted \cite{cimpanu_2020}. 
The attack targeted vulnerable computers in the distribution networks, that control water flow and wastewater treatment, as well as programmable logic controllers (PLCs) that operate valves in a number of locations. 
The systems recorded faulty data, pumps went out of control, and the attackers took over the operation system at one station.

CPSs are characterized by complex physical phenomena, typically modeled by means of approximation algorithms which are very computationally taxing and often not accurate enough. State-of-the-art approaches for securing CPS rely heavily on physics-based models of the physical side \cite{rai2020driven}. However, carefully engineered attacks through the cyber layer have been demonstrated to cause significant system failures while bypassing any physics-based intrusion detection system. Methods able to merge data-driven learning with physics-based models can build algorithms able to significantly enhance the security of CPS to cyberattacks by concurrently leveraging both cyber-layer and physical-layer information \cite{8743447}.
An emerging category of neural networks, physics-informed neural networks (PINNs), have demonstrated great ability in data-driven training to understand the physical limitations of a given scenario and perform very accurate approximations when regarding physical laws that are described by partial differential equations. This results to fast data-driven inference of partial differential equation solutions. 

In this work, we examine the feasibility of using PINNs in a test case of a smart water distribution system in order to mitigate the effect of sensor-based attacks. We examine a proof-of-concept scenario of a controller in charge of dictating a pump's operation based on measurements from sensors placed within a pipe. In order to deal with the attacked (data integrity) or removed (availability) measurements while retaining system observability, a trained PINN acts as a virtual sensor to prevent the controller from altering the pump's operating conditions and mislead state estimation processes. We examine the applicability of the approach while discussing the challenges and future applications. 


The rest of the paper is organized as follows. Section \ref{s:Background} provides background on PINNs. 
Section \ref{s:Methodology} presents our threat model and proposed approach. The setup and results are presented in Section \ref{s:Experimental Setup}. Challenges and future work are discussed in Section \ref{s:Challenges and Future Work}.

\section{Physics-Informed Neural Networks (PINNs)} \label{s:Background}
PINNs are neural networks trained to solve supervised learning tasks while respecting given law of physics, in the form of general nonlinear partial differential equations \cite{raissi2019physics}. Therefore, they can be integrated in complex physical systems and provide surrogate models that naturally encode the system's underlying physical laws. 
A set of partial differential equations that PINNs can be trained to approximate are the Navier-Stokes equations \cite{girault1979finite}. These equations describe the motion of viscous fluid substances which encompasses many phenomena 
such as air flow around a wing and water flow in a pipe. Our use case is focused in the two-dimensional adaptation of the equations, given by: 
\begin{equation}
    \label{eq:continuity}
    \frac{\partial u}{\partial x} + \frac{\partial v}{\partial y} = 0
\end{equation}
\begin{equation}\label{eq:xvelocity}
    \frac{\partial u}{\partial t} + u\frac{\partial u}{\partial x} + v\frac{\partial u}{\partial y} = -\frac{1}{\rho}\frac{\partial p}{\partial x} + \nu(\frac{\partial^2 u}{\partial x^2} + \frac{\partial^2 u}{\partial y^2})
\end{equation}
\begin{equation}\label{eq:yvelocity}
    \frac{\partial v}{\partial t} + u\frac{\partial v}{\partial x} + v\frac{\partial v}{\partial y} = -\frac{1}{\rho}\frac{\partial p}{\partial y} + \nu(\frac{\partial^2 v}{\partial x^2} + \frac{\partial^2 v}{\partial y^2})
\end{equation}
where $u(x,y,t)$ is the $x$-component of the velocity field, $v(x,y,t)$ the $y$-component, and $p(x,y,t)$ the pressure. The kinematic viscosity of the fluid is denoted by $\nu$ and its density by $\rho$.
Eq. \eqref{eq:continuity} is the continuity equation, derived from the principle of conservation of mass. Eqs. \eqref{eq:xvelocity} -- \eqref{eq:yvelocity}, represent the relationship of $U$ and $V$ velocity, respectively, with pressure $p$, time $t$, and spatial coordinates $(x,y)$.

PINNs trained for Navier-Stokes equations can be leveraged for fast prediction of the system's state at given time instants, since velocity and pressure represent the physical properties of a system's environment. Utilizing this ability, the PINN can act as a surrogate sensor in case of sensor compromise during cyberattacks or malfunctions.


\begin{figure}[t]
    \includegraphics[width=\linewidth]{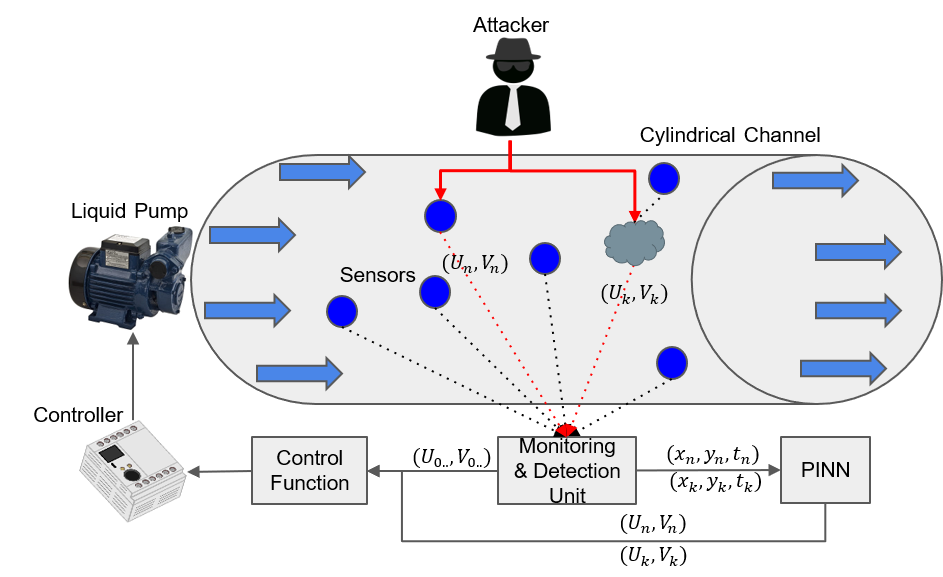}
    \caption{PINN-enhanced smart water distribution network: A liquid pump's operation is dictated by the decisions of a control function. The control routine 
    collects data from remote sensors that reports the cylindrical channel's state. An attacker tries to indirectly influence the operation of the pump by altering the data sent to the controller from the sensor by either complete sensor compromise or man-in-the-middle interference.}
    \label{fig:overview}
    \vspace{-0.2in}
\end{figure}
\section{Methodology} \label{s:Methodology}

A smart water distribution network, as a CPS, tightly integrates sensor measurements that dictate the behavior of system controllers and actuators. The decision processes within CPSs often involve information feedback loops that control software and hardware systems to take informed actions towards specific efficient and stable operation objectives. Hence, for large-scale critical CPS, e.g., water supply networks, sensors are placed in remote, often exposed, locations and communicate with the system's control and estimation functions through wireless networks. 
These remote sensors constitute an alluring target for potential cyberattacks in the form of \emph{data availability} and \emph{integrity attacks}. Availability attacks (e.g., denial-of-service, SYN flood attacks, etc.) can render the water distribution network unobservable, i.e., unable to estimate system states due to insufficient information gathered from the collected measurement data. Similarly, integrity attacks (e.g., false data injection attacks \cite{mo2010false,8894484}) can manipulate and falsify the collected sensor measurements, overall altering state estimation outcomes. 
Such \emph{availability}-based attacks as well as \emph{data integrity} \emph{attacks} targeting \emph{observability} can lead CPS operators to be oblivious of the overall system state leading to incorrect control actions and, thus, causing harmful cascading events\cite{sayghe2020adversarial}. Leveraging the underlying physics that characterize the CPS with neural networks can aid in state estimation.


In this section, we explain our approach which aims to provide means of mitigation to such attacks. Our proof-of-concept scenario consists of an indicative part of a smart water distribution system in which a pump pushes liquid through a pipe and aims to maintain a constant flow by adjusting its operating conditions according to measurements captured by sensors placed in the pipe. An illustration of the considered setup can be seen in Fig. \ref{fig:overview}.

\subsection{Threat Model} \label{ss: Threat Model}
In this work, we consider a case of an incompressible fluid flowing past a circular cylinder. We assume non-dimensional free stream velocity, a cylinder with constant diameter, and that the fluid is uniform in density across the cylindrical channel. Sensor are placed inside the channel and record the fluid's $U$ and $V$ velocity. The sensors report the measurements to a control algorithm (e.g., state estimation) in order for the latter to take appropriate actions via controllers and actuators for the liquid pump operation. An attacker, aware of this feedback loop, tries to interfere with the reported measurements to indirectly mislead the controller and steer the system towards amplifying the error in the estimation, and thereby leading to non-convergence or even towards a possibly unstable state.

\subsection{System Observability}
 
In a CPS, physical processes are typically monitored in order to provide control routines with the appropriate state information. This information acts as external stimuli that depicts a clear picture of what is happening at different parts of the system at any given moment, and helps controllers to dynamically react to the current conditions. 


\begin{figure}[t]
    \centering
    \includegraphics[width=\linewidth]{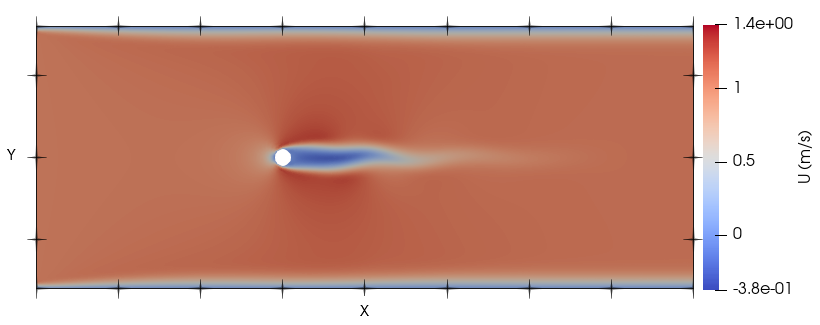}
    \caption{The $x$-component ($U$) velocity field representation of the whole channel as depicted by the graphical representation tool \textit{ParaView}. Liquid flows from left to right with top and bottom boundaries acting as solid wall surfaces. A small vortex is created in the channel due to the sensor and points directly behind it move slower in the $x$-direction.}
    \label{fig:velocity}
    \vspace{-0.2in}
\end{figure}

\begin{definition}[Observability]\label{def:observability}
A system is said to be observable if, for any possible evolution of state and control vectors, the current state can be estimated using only the information from system outputs. For an input-free time-invariant continuous system:
\begin{equation}\label{eq:x}
        \dot{x}(t) = \mathbf{A}x(t), \hspace{5mm}
        x(t_0) = x_0 = unknown
\end{equation}
with measurements:
\begin{equation}\label{eq:y}
    y(t) = \mathbf{C}x(t)
\end{equation}
where $x(t)\in\Re^n$, $y(t)\in\Re^p$, $\mathbf{A}\in\Re^{n\times n}$, and $\mathbf{C}\in\Re^{p\times n}$, the system is observable if the observability matrix $\mathcal{O}=[C, CA,\cdots, CA^{n-1}]^T \in \mathbb{R}^{n p \times n}$ has full column rank (i.e., $\operatorname{rank}(\mathcal{C})=n$). In our work, we consider a single sensor ($n=1$) placed in the cylindrical channel which collects two measurements ($p=2$): the $x$ and $y$ components of the liquid's velocity, as seen in Fig. \ref{fig:velocity}.
\end{definition}

In order to derive the state of the system at a given time using Eqs. \eqref{eq:x} -- \eqref{eq:y}, $x_0$ has to be determined. Since the $n$-dimensional vector $x(0)$ has $n$ unknown components, $n$ measurements are required to sufficiently determine $x_0$. If $n$ derivatives of the continuous time measurements are used, the observability matrix $\mathcal{O}$ can be formed. Following Definition \ref{def:observability}, for the system to be considered observable, the rows of $\mathcal{O}$ have to be linearly independent. However, if an attacker is able to tamper the sensor measurements, the modified $\mathcal{O}$ contents can render the system unobservable. Thus, compromising the measurements can indirectly affect the pump's operation.

\subsection{Attack Mitigation Strategy}


\begin{figure}[t]
    \centering
    \includegraphics[scale=0.5]{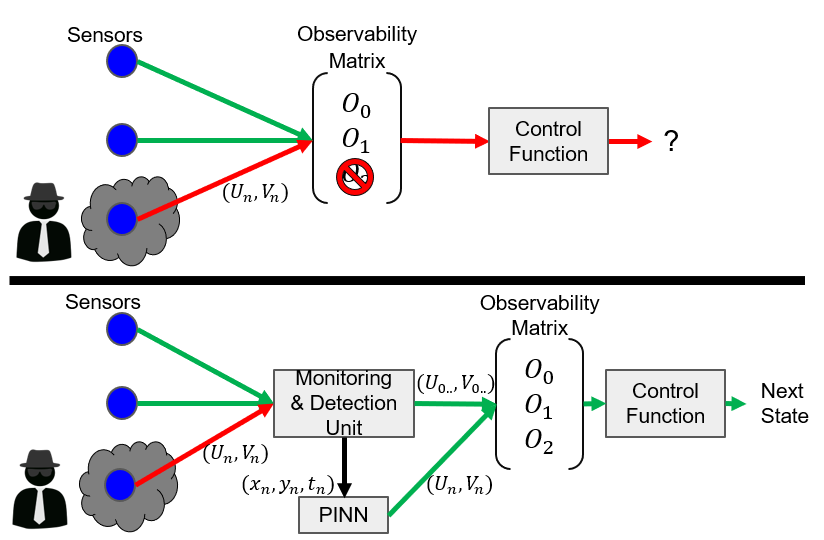}
    \caption{\textbf{Top:} An attacker compromises a sensor, making the system unobservable and affecting the functional operation of CPS control routines. \textbf{Bottom:} The detection of falsified data follows the replacement with PINN-generated values to retain system observability.}
    \label{fig:methodology}
    \vspace{-0.2in}
\end{figure}

Considering the above threat model and observations, typical control routines in CPS such as state estimators include monitoring and detection units (MDUs), able to identify corrupted sensing data \cite{ahmed2017model, 8990004}. 
MDUs contain detection algorithms which can remove faulty or malicious data so that they will not affect the control estimation routines \cite{7867816, 8894484}. However, it is necessary that the underlying physical system state to be computed from existing measurements (system observability). In order to address this issue of removing the detected and contaminated data, and thus, keep the system observable, a trained PINN can be activated and act as a temporary surrogate data provider to the control function (Fig. \ref{fig:methodology}). The MDU is aware of the stationary sensors' $x$ and $y$ coordinates and thus forwards the compromised sensor's coordinates to the PINN, alongside with the current time instance. Then, the PINN calculates the corresponding $U$ and $V$ values under normal operating conditions, to fill in the gap due to the discarded sensor measurements.

PINNs' accuracy in recreating the dynamic environment of a physical system by taking into account the bounds of physical laws can act as reliable sources of very accurate approximate information. By replacing the compromised sensor during the attack timeframe with a ``virtual PINN-enabled sensor'', the system can maintain its observability and continue its normal operation. The utilization of PINNs, in this scenario as a fail-safe mechanism, can give enough time to the system operators to thwart the attack or even replace a malfunctioning sensor without having to shut down the system or lose control of it. 


\section{Experimental Setup \& Results} \label{s:Experimental Setup}
To validate and evaluate PINNs ability to predict a system's dynamic environment for the case of liquid flow in a pipe, we first create an appropriate dataset following the example in \cite{raissi1711physics}. In particular, we create a cylindrical channel using \textit{gMsh} by creating a rectangular mesh geometry, discretazing space in triangles for the computational solver \cite{geuzaine2009gmsh}. 
We assume a uniform free stream velocity at the left boundary, a zero pressure outflow condition imposed at the right boundary, and set the top and bottom boundaries as walls. The channel encloses a domain of size $[-15,25]\times[-8,8]$.


\begin{figure}[t]
    \centering
    \includegraphics[width=\linewidth]{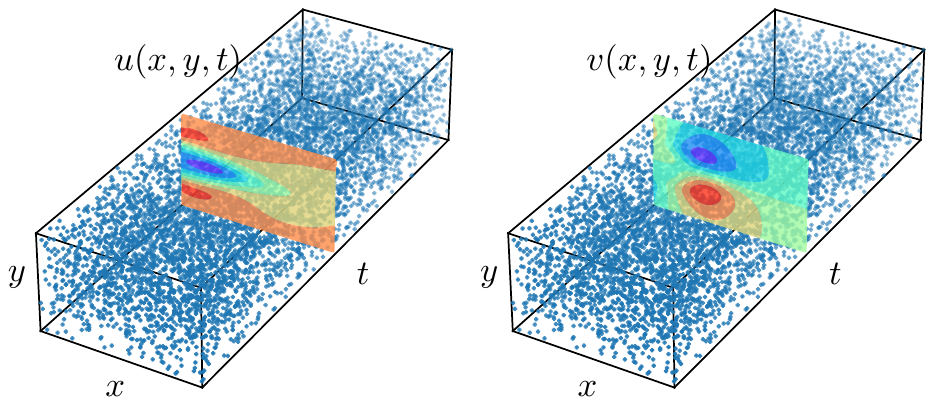}
    \caption{A snapshot representation of the velocity $(U,V)$ field predicted by the trained PINN as a slice of the $(x,y,t)$ testing dataset. The scattered blue dots represent the datapoints generated by the finite-element solver of the Navier-Stokes equation, included in \textit{NekTar++}.}
    \label{fig:snapshot}
    \vspace{-0.1in}
\end{figure}

For creation of a high-resolution dataset, we utilize the spectral/\textit{hp}-element solver \textit{NekTar++} \cite{cantwell2015nektar++}. In order to use the built-in Navier-Stokes solver in \textit{NekTar++}, we create the appropriate configuration .XML files. For the sake of simplicity, we define Reynolds number $Re = 100$, kinematic viscosity $\nu = 0.01$, free stream velocity $u_\infty = 1$, and a total experiment duration of $2000$ steps. A data-snapshot of the system is taken every $10$ steps. A snapshot of the component velocity field of the resulting simulation is depicted in Fig. \ref{fig:snapshot}.

\begin{table}[t]
    \centering
    \caption{PINN Accuracy in Velocity Prediction}
    \label{tab:results}
    \begin{tabular}{||c||c||}
    \hline \hline
    Velocity Component & Accuracy (\%) \\ \hline
    $U$     &  99.4511313 \\ \hline 
    $V$     &  95.712483 \\ 
    \hline \hline
    \end{tabular}
    \vspace{-0.2in}
\end{table}

The resulting solution to the Navier-Stokes equations is split to training and testing data with a ratio of 70\% and 30\%, respectively. Building on top of the PINN code in \cite{maziarraissi}, we have altered its operation mode from identifying Navier-Stokes equations into inferring them. The neural network consists of an input layer $(x,y,t)$, $7$ hidden layers with $20$ neurons each, and an output layer $(U,V)$. Each layer has a hyperbolic tangent activation function. Our training dataset only represents a small area of the channel, specifically the area $[1,8]\times[-2,2]$, in order to better demonstrate the network's ability to generalize. The results of the testing session are given in Table \ref{tab:results}. The testing set, i.e., 30\% of the generated dataset, consists of 2087 spatial coordinates, in 200 temporal points each. The test's accuracy is normalized and averaged in order to provide a single metric for $U$ and $V$ velocity predictions. Specifically, the trained PINN is able to successfully approximate the $x$-coordinate and $y$-coordinate of velocity, $U$ and $V$, of the points tested with
a deviation of $\approx0.55\%$ and $\approx4.3\%$ from the actual ground-truth values, respectively.

\vspace{-0.4em}
\section{Challenges and Future Work} \label{s:Challenges and Future Work}

In this paper, we demonstrate the applicability of PINN as a mitigation mechanism against observability cyberattacks in CPSs. We show that a PINN can be very accurate with minimal training and a small dataset. This way, complex physical processes can be accurately characterized without the computational complexity and effort of traditional approximation methods. In order to further validate this specific idea, it is important to determine how long the accuracy of the PINN can remain high, hence rendering it an effective mitigation mechanism which enhances the availability of a system. In addition, investigating the accuracy of PINNs trained using real datasets derived from real-world water distribution facilities and digital twin testbeds is an immediate future task, which will further validate the effectiveness of PINNs to model the physical dynamics of the system under consideration.

A few challenges are still at large, such as the mitigation mechanism of the current work, which assumes a detection mechanism against falsified data. Furthermore, the current approach cannot distinguish between faulty sensors and cyberattacks. It relies on the detection mechanism's ability to do that. As the detector is a crucial component of the overall system, in our future work we will investigate how PINN generated data can be used to detect measurement anomalies caused by data integrity and availability attacks. The potential to utilize PINNs in various CPS domains for anomaly detection and state estimation stems from their accuracy. PINNs could be utilized as ground truth when comparing the system's current state against the PINN-predicted one, to distinguish anomalous behavior beyond problematic sensor measurements. For example, the Navier-Stokes equations' solutions rely on the liquid's density. Therefore, a change in density, e.g. contamination of the water, could be detected. 

\vspace{-0.3em}
\section*{Acknowledgment}\label{s:Acknowledgement}
This work has been supported by the European Union’s Horizon 2020 research and innovation programme under grant agreement No 739551 (KIOS CoE) and from the Government of the Republic of Cyprus through the Directorate General for European Programmes, Coordination and Development.
\vspace{-0.5em}

\bibliographystyle{IEEEtran}
\bibliography{sources}




\end{document}